\documentclass[]{aa}  

\usepackage{graphicx}
\usepackage[varg]{txfonts}
\usepackage[english]{babel}
\usepackage{natbib}
\usepackage{amssymb}  
\usepackage{amsmath}
\usepackage{footnote}

\usepackage[colorlinks,citecolor=blue,linkcolor=blue]{hyperref}
\begin{document}

   \title{Hydrodynamic Modeling of Accretion Impacts in Classical
   T Tauri Stars: Radiative Heating of the Pre-shock Plasma}

   \author{G. Costa
          \inst{1,3},
          S. Orlando
          \inst{2},
          G. Peres
          \inst{1,2},
          C. Argiroffi
          \inst{1,2}
          R.Bonito
          \inst{1,2}
          }
          
   \institute{Dipartimento di Fisica \& Chimica, Universit\`a di Palermo,
              Piazza del Parlamento 1, 90134 Palermo, Italy
\and
             INAF-Osservatorio Astronomico di Palermo, Piazza del Parlamento 1, 90134 Palermo, Italy   		
   		\and
	          Astrophysics sector, SISSA - International School for Advanced Studies, Via Bonomea 265, 34136, Trieste, Italy}

   \date{Received ; accepted }

 
  \abstract
   {It is generally accepted that, in Classical T Tauri Stars, 
   the plasma from the circumstellar disc accretes onto the stellar
   surface with free fall velocity, and the impact generates a shock.
   The impact region is expected to contribute to emission in
   different spectral bands; many studies have confirmed that the 
   X-rays arise from the post-shock plasma but, otherwise, there are no 
   studies in the literature investigating the origin of the observed Ultra-Violet
   emission which is apparently correlated to accretion.}
   {We investigated the effect of radiative heating of the
   infalling material by the post-shock plasma at the base of the
   accretion stream with the aim to identify in which region
   a significant part of the UV emission originates.}
   {We developed a one dimensional hydrodynamic model
   describing the impact of an accretion stream onto the
   stellar surface; the model takes into account the gravity, the
   radiative cooling of an optically thin plasma, the thermal
   conduction, and the heating due to absorption of X-ray radiation.
   The latter term represents the heating of the infalling plasma 
   due to the absorption of X-rays emitted from the post-shock region.}
   {We found that the radiative heating of the pre-shock
   plasma plays a non-negligible role in the accretion phenomenon.
   In particular, the dense and cold plasma of the pre-shock accretion
   column is gradually heated up to few $10^5$\;K due to irradiation
   of X-rays arising from the shocked plasma at the impact
   region.  This heating mechanism does not affect significantly
   the dynamics of the post-shock plasma. On the other hand, a
   region of radiatively heated gas (that we consider a precursor)
   forms in the unshocked accretion column and contributes
   significantly to UV emission. Our model naturally reproduces the
   luminosity of UV emission lines correlated to accretion
   and shows that most of the UV emission originates from the precursor.}
   {}


\titlerunning{Hydrodynamic Modeling of Accretion Impacts in Classical
   T Tauri Stars}
\authorrunning{G. Costa et al.}
\maketitle

%
\section{Introduction}

Classical T Tauri Stars (CTTS) are young (age $\lesssim 10\;$My),
low-mass ($\rm M_* \simeq  1 \; M_\odot$) stars surrounded by
circumstellar disks and subject to accretion/outflow phenomena.
These young stars radiate at all wavelength bands, and in
particular, are strong UV and X-ray emitters.  This evidence, and
complementary theoretical studies, support the widely accepted
magneto-spheric accretion scenario \citep[e.g. see review
by][]{Bouvier2007prpl.conf..479B}.  The stellar magnetic field
drives the infalling plasma from the truncation radius of the disk
(R$_T\sim3-10\;$R$_*$) onto the stellar surface, often at high
latitude \citep{Koenigl1991}.  The gravitational field accelerates
the plasma up to supersonic velocities ($\sim 300-600\;$km/s) and
then, the impact onto stellar surface generates a strong shock that
heats the plasma up to temperature of a few MK
\citep{Calvet1998ApJ...509..802C}.

In the last decade, high spectral resolution
X-ray observations have revealed that the X-ray emission 
from these objects originates from two components: the dense 
($n_e>10^{11}\; $cm$^{-3}$) and cold ($T \sim 2-5\;$MK) plasma 
heated by the accretion shock, that dominates the soft X-ray
emission ($E\leq 1\;$keV) \citep{Kastner2002,Schmitt2005,Gunther2007}, 
and the hotter and less dense 
($n_e \leq 10^{10} \; $cm$^{-3}$) coronal plasma, with typical
temperature of $10-20\;$MK,  that dominates the harder X-ray emission
($E>1\; $keV) \citep{Brickhouse2010ApJ...710.1835B,Argiroffi2011}. 
Observations of CTTSs in the UV band have shown emission lines
of plasma with temperature $T \sim 10^5 \rm \; K$ clearly related
with accretion phenomena, because of the observed Doppler
shift \citep{Ardila2002,Gunther2008,Ardila2013ApJS..207....1A}, and
because of much higher luminosity than those observed in young
stars without accretion \citep{Johns-Krull2000}. Some
authors have suggested that the UV emission may originate from a
region in the accretion flow which is immediately before the
accretion shock at the base of the column
\citep{Herczeg2002,Ardila2013ApJS..207....1A}.

Several theoretical studies have provided important support to the 
magnetospheric accretion scenario and crucial knowledge about the accretion 
shock phenomena.
One dimensional (1-D) hydrodynamic (HD) models
\citep[e.g][]{Sacco2008,Sacco2010} studied the dynamics and the
structure of the impact region of a continuous accretion inflow
onto the stellar surface.  These models take into account many
important physical effects: the thermal conduction, the radiative
cooling from optically thin plasma, the gravity stratification, and
a detailed description of the stellar chromosphere. In
the case of MP Mus, a well studied CTTS, these models were able to
reproduce the main features of its high resolution X-ray spectrum
collected with RGS (Reflection Grating Spectrometers) on board the
XMM-Newton satellite \citep{Argiroffi2007A&A...465L...5A}, thus
showing an excellent agreement between models and observations
\citep{Sacco2008}. These models assume that the magnetic field
dominates the dynamics and the energetics of the downflowing plasma
(i.e. $\beta \ll 1$, where $\beta =$ gas pressure/magnetic pressure).
\citet{Sacco2010} have suggested that the UV emission may arise from
the surrounding chromospheric plasma, because of the absorption of X-rays
originating in the post-shock region rooted in the chromosphere,
or maybe in the accretion stream itself. Studies of
\citet{Sacco2010,Reale2013,Bonito2014ApJ...795L..34B} have
shown that the absorption of X-rays by optically thick material
distributed around the post-shock plasma can be significant and
may play a relevant role in determining the emission emerging 
from the impact region. 

Further studies were performed with 2-D magnetohydrodynamic
(MHD) models of the accretion stream impact
\citep{Orlando2010,Orlando2013A&A...559A.127O,Matsakos2013}. These
2-D simulations showed that the structure and dynamics of
the post-shock plasma strongly depend on the strength and configuration
of the ambient magnetic field. If $\beta < 1$ then the accretion
stream is divided into elementary fibrils, each with dynamic and
evolution independent from those of the other, that can be described by 1-D
simulations \citep{Sacco2008,Sacco2010}. These 2-D models showed
that, if $\beta$ is close to or higher than 1, the accretion stream
may strongly perturb the nearby stellar atmosphere and, in case of
$\beta \gg 1$, the shock may generate some plasma leaks into the
surrounding corona. Studies of \citet{Brickhouse2010ApJ...710.1835B}
support the idea that the accretion shock may perturb also the
stellar corona.

Thanks to the high quality observations and the highly refined HD
and MHD models our knowledge of accretion phenomena in CTTSs is
strongly growing in this decade. However there are several aspects
that still remain unclear.  There are no theoretical models from
which predictions of both UV and X-ray emission are derived at the
same time and compared with observations. Many studies have shown
that the X-ray emission arises from the shock heated plasma.
However, there is no certainty on where the UV emission
comes from.  Theoretical models proved that absorption of X-rays
in the accretion-shock region is significant
\citep{Bonito2014ApJ...795L..34B} but the consequences of
local heating due to this absorption are still unexplored.
Investigating these aspects could probably help in understanding
one of the most debated issues: the evidence that the mass accretion
rates inferred from X-ray observations are systematically lower
(even by one or more orders of magnitude) than those derived from
NIR/optical/UV observations \citep{Curran2011A&A...526A.104C}.

In this paper, we investigated the radiative heating of the
infalling material by the post-shock plasma at the base of the
accretion stream. The main aims are to investigate: (1) how the
radiative heating affects the dynamics and energetics of the
accretion shock; (2) if, and to what extent, the UV and X-ray
emission arising from the shocked plasma may heat the cold pre-shock
material of the accretion column; (3) the source region(s) 
of the observed UV emission; 
(4) if the radiative heating of the pre-shock
material may reproduce simultaneously the observed UV and X-ray
emission. To answer these questions, we assumed $\beta\ll 1$ and
performed 1D HD simulations of accretion impacts \citep[as
done in][]{Sacco2008} but including the effects of radiative
heating on the dynamics and energetics of the system. In Sect.
\ref{sec.2} we describe our model and numerical setup; in Sect.
\ref{sec.3} we describe our results and in Sect. \ref{sec.5}
we draw our conclusions.


\section{The model}
\label{sec.2}

\subsection{The hydrodynamic model}
\label{sec.2.1}

Following \cite{Sacco2010}, we developed a 1-D HD model describing
the impact of an accretion stream onto the surface of a CTTS. We
assumed that accretion occurs along a magnetic flux tube linking
the circumstellar disk to the star and the plasma moves and transports
energy exclusively along the magnetic field lines ($\beta \ll 1$).
Thus our model describes the impact of one fibril of the accretion
stream onto the stellar surface.

\begin{figure}
  \resizebox{\hsize}{!}{\includegraphics{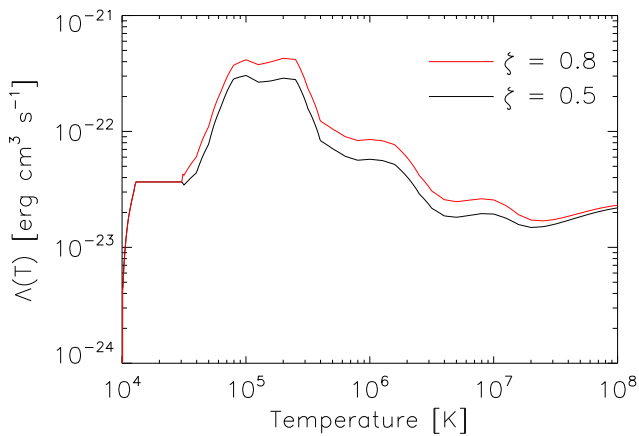}}
  \caption{Radiative losses for an optically thin plasma from the
  CHIANTI atomic database
  \citep{Dere1997A&AS..125..149D,Landi2013ApJ...763...86L}, assuming
  the metal abundance of 0.5 (black line) and 0.8 (red line) of the solar value.}
  \label{fig:Perd.Rad}
\end{figure}

The impact is modelled by solving the time-dependent HD
equations of conservation of mass, momentum, and energy. The model
takes into account the gravity, the thermal conduction, the
radiative losses from an optically thin plasma, and the heating
of the accretion stream plasma due to the absorption of the radiation
which arises from the post-shock zone

\begin{equation}
\centering
	\frac{\partial \rho}{\partial t} + \frac{\partial \rho v}{\partial z} = 0 ,
	\label{eq:cons.mass}
\end{equation}
\begin{equation}
\centering
	\frac{\partial \rho v}{\partial t} + \frac{\partial (P + \rho v^2)}{\partial z} =	\rho g,
	\label{eq:cons.mom}
\end{equation}
\begin{equation}
\centering	
	\frac{\partial E}{\partial t} + \frac{\partial (E + P)v}{\partial z} = \rho u g -\frac{\partial F_c}{\partial z} -n^2_e \Lambda (T) + H(z),
	\label{eq:cons.ene}
\end{equation}
\begin{equation}
\epsilon = 3 \,k_b T/ \mu m_H, \qquad E=\frac{1}{2} \rho v^2 +\rho \epsilon,
\end{equation}

\noindent
where $\rho = \mu \, n_H \, m_H$ is the mass density; $\mu = 1.277$ 
(1.284) is the mean atomic mass, assuming metal abundances
 of 0.5 (0.8) of the solar value; 
$m_H$ is the mass of the hydrogen atom; $n_H$
is the hydrogen number density; $t$ is the time; $z$ is the
 coordinate
along magnetic field; $v$ is the plasma velocity along $z$; $P$ is
the thermal pressure; $g(z)$ is the gravity of a star with a mass
$ M_* = 1.2\, M_\odot$ and a radius $R_*= 1.3\,R_\odot$, values of the
CTTS MP Mus; $E$ is the total energy per unit volume; $F_c$ is the
conductive flux; $n_e$ is the electron number density; $\Lambda
(T)$ is the radiative losses function per unit emission measure
from an optically thin plasma (see Fig. \ref{fig:Perd.Rad}); $H(z)$
is the heating function (see Sect.~\ref{sec.2.2}); $\epsilon$ is
the internal energy per unit mass; $k_b$ is the Boltzmann constant;
$T$ is the plasma temperature. We used the ideal gas law,
$P = (\gamma - 1) \rho \epsilon$, where $\gamma = 5/3$ is the ratio
between the specific heats.

The thermal conduction includes the classical and
the saturated regime, and a transition between the two given by

\begin{equation}
\centering
	F_c = \Bigl(\frac{1}{q_{spi}} + \frac{1}{q_{sat}} \Bigr)^{-1},
	\label{eq:Cond.Flux}
\end{equation}
where
\begin{equation}
\centering
	q_{spi} = -\kappa \partial T/ \partial z \approx - 9.2 \times 10^{-7} T^{5/2} \partial T/ \partial z
\end{equation}
\begin{equation}
	q_{sat} = - sign( \partial T/ \partial z) 5 \phi \rho	c^3_s
\end{equation}
are, respectively, the classical conductive flux 
\citep{Spitzer1962pfig.book.....S} and the saturated flux 
\citep{Cowie1977}, where $\kappa$ is the thermal conduction coefficient, $\phi \leq 1$ \citep[and references 
therein]{Borkowski1989,Fadeyev2002}, and $c_s$ is the 
isothermal sound speed.

The radiative losses per unit emission measure are derived
with the PINTofALE spectral code \citep[Package for Interactive
Analysis of Line Emission,][]{Kashyap2000BASI...28..475K} adopting
the CHIANTI atomic database \citep{Dere1997A&AS..125..149D,
Landi2013ApJ...763...86L}. The radiative losses depend on the metal
abundance, $\zeta$, in the range of temperature between $10^4 < T
< 10^7 \;$ K, because the radiative emission is dominated by lines
of heavy ions (see Fig. \ref{fig:Perd.Rad}). Here we assumed
heavy elements abundances of 0.5 and 0.8 the solar values, 
in agreement
with X-ray observations of CTTSs by \citet{Telleschi2007}. 

We noted that, in our simulations, the plasma in the hot slab at the
base of the accretion column can be considered to be optically
thin, because of its density, temperature, and dimension. The
chromosphere is known to be optically thick and we accounted
for this by setting the radiative losses equal to zero there.
We assumed that the pre-shock material of the accretion
stream is cold ($T=10^{3}$ K) when it starts its falling. Given
its density and temperature, the infalling material of the accretion
stream cannot be considered to be, in general, optically thin. As
a consequence the radiative losses adopted in our model are not
appropriate to describe the material in the unshocked stream. To
account for this, we explored two limiting cases by adopting
two complementary approaches: 1) we assumed that the stream material
is completely opaque to its radiation by setting the radiative
losses equal to zero in the unshocked stream; 2) we assumed that
the stream material is optically thin and any effect of self-absorption
of radiation is negligible. In
such a way, the radiative cooling of the plasma in the pre-shock
stream is neglected in the first case and overestimated in the
second. We discuss the limits of these assumptions in Sect.
\ref{sec.2.4}.

The model is implemented with PLUTO
\citep{Mignone2007ApJS..170..228M}, a modular Godunov-type code for
astrophysical plasmas, designed to use massively parallel computers
using the Message-Passing Interface (MPI) for interprocessor
communications. The HD equations are solved using the linearized
Roe Riemann solver based on the characteristic decomposition
of the Roe matrix \citep{Roe1981}. PLUTO includes a module to
calculate the optically thin radiative losses at the temperature
of interest, using a table-lookup/interpolation method. The thermal
conduction is computed with explicit time-stepping schemes for
parabolic problems.

The star and the accretion flow parameters of the simulations have
been chosen to reproduce accretion streams as inferred from X-ray
observations of the CTTS MP Mus \citep{Argiroffi2007A&A...465L...5A},
and expected to produce detectable X-ray emission as described by
\cite{Sacco2010}. The computational domain extends from $0.98\;
R_*$ to $1.7\; R_*$ (where $R_* = 1.3\,R_{\odot}$); the transition
region is at $z=1\;R_*$. It is worth noting that the domain
adopted here is much larger than the height of the post-shock slab
expected to develop at the base of the accretion column. This choice
allowed us to make the effects of the upper boundary (at
$z=1.7\;R_*$) negligible for the evolution of the slab. On the other hand, the
adopted domain is smaller than the length expected for a stream linking
the star to its disk. In fact our study focusses on the impact
region close to the stellar surface and our 1D model cannot describe
properly the complex geometry expected for the whole stream.
We divided the domain in two regions. The first, with the highest
resolution (about $1.2 \times 10^6\;$ cm), consists of an uniform
grid with 9000 points, which extends from $0.98\; R_*$ to $1.1\;
R_*$. The slab is expected to evolve in this region. The
second region consists of a non-uniform stretched grid with 11000
points, which extends from $1.1\; R_*$ to $1.7\; R_*$. In this
latter case, the grid spacing grows slowly from $\rm \sim 1.2\times
10^6 \;cm$ to $\rm \sim 1.3 \times 10^7 \;cm$ at the end of the
domain. We performed a grid convergence study to check how
the numerical solution depends on the spatial resolution;in particular,
we performed the simulation of our reference case on more successively 
finer grids until the solution does not change by more than a few percent. 
We found that the adopted grid represents the best compromise between 
accuracy and computational cost.

The initial condition describes the accretion stream just before
the impact onto the stellar chromosphere. Initially the
chromosphere extends from the lower boundary ($z=0.98\;R_*$) up to
$z = 1\;R_*$ and the stream fills the rest of the mesh from $z =
1\;R_*$ up to the upper boundary ($z=1.7\;R_*$); the model does not
describe the stellar corona between the chromosphere and the stream.
The lower boundary ($z=0.98\;R_*$) is set assuming an
isothermal chromosphere with $T_{ch}=10^4 \rm\;K$, in hydrostatic
equilibrium with a minimum density $n_{ch} = 7\times 10^{10}\rm \;
cm^{-3}$ at $z=1\rm \; R_*$ (namely the top of the chromosphere).
We adopted an isothermal chromosphere for simplicity: models with
a more accurate chromosphere did not show significant differences
in the results \citep{Sacco2008,Sacco2010}. From here the '$post$' subscript indicates the post-shock quantities,
while the '$acc$' subscript indicates the accretion stream quantities. The upper
boundary ($z=1.7\;R_*$) is set assuming a continuous inflow of
plasma with fixed values of temperature, $T_{acc}=10^3\; $K, and
velocity, $v_{acc}= -500 \; $km/s. We explored two possible values
of stream density, either $n_{acc} \rm = 10^{11} \;cm^{-3}$ or
 $n_{acc} \rm = 5 \times 10^{11} \; cm^{-3}$, and two possible 
 values of metal
abundance, either $\zeta = 0.5$ or $\zeta = 0.8$. Table
\ref{tab:par.mod} summarizes the parameters characterizing the
models explored: the accretion stream density, $n_{acc}$,the 
abundance with respect to the solar one,$\zeta$, temperature,
$T_{post}$, density, $n_{post}$, characteristic quasi-periodic
time, $P_o$, maximum post-shock region length, $L_{post}$,
precursor region length, $L_{prec}$, and maximum precursor
temperature, $T_{prec}$ (the precursor will be defined in Sec.\ref{sec.3.2}); in the table note, the accretion velocity, 
$v_{acc}$, and the post-shock plasma velocity, $v_{post}$. 
The parameters explored are those
required to match the soft X-ray emission of many CTTSs (e.g. MP
Mus, \citealt{Argiroffi2007A&A...465L...5A, Sacco2010}). In addition
these models allowed us to compare our results with those discussed
in the literature \citep{Sacco2008,Orlando2010}.  Our simulations
follow the evolution of the system for $\sim 3.6 \;$ks.

\begin{table*}
\caption{Parameters characterizing the models.} 
\label{tab:par.mod} 
\centering
\begin{tabular}{ccccccccc}
\hline\hline 
Model & $n_{acc}$ & $\zeta$ & $T_{post}$ & $n_{post}$ & $P_o$ & $L_{post}$ & $L_{prec}$ & $T_{prec}$ \\ 
name & [cm$^{-3}$] & & [K] & [cm$^{-3}$] & [s] & [cm] & [cm] & [K] \\
\hline 
D1e11-A05 & $10^{11}$ & 0.5 & $ 6.6\times 10^6$ & $6-7\times 10^{11}$ & $ 550$ & $3.9\times 10^9$ & $-$ & $-$ \\
D1e11-A08 & $10^{11}$ & 0.8 & $ 6.3\times 10^6$ & $4-5\times 10^{11}$ & $ 510 $ & $2.3\times 10^9$ & $-$ & $-$ \\
D5e11-A05 & $5 \times 10^{11}$ & 0.5 & $ 6.5\times 10^6$ & $3-4\times 10^{12}$ & $ 100$ & $6.5\times 10^8$ & $-$ & $-$ \\
D5e11-A08 & $5 \times 10^{11}$ & 0.8 & $ 6.1\times 10^6$ & $2-3\times 10^{12}$ & $ 80$ & $2.8\times 10^8$ & $-$ & $-$ \\
D1e11-A05-RT & $10^{11}$ & 0.5 & $ 6.1\times 10^6$ & $5-6\times 10^{11}$ & $ 610$ & $4.6\times 10^9$ & $2.53 \times 10^{10}$ & $ 1.4 \times 10^6$ \\
D1e11-A08-RT & $10^{11}$ & 0.8 & $ 6.7\times 10^6$ & $4-5\times 10^{11}$ & $ 530$ & $3.0\times 10^9$ & $1.77 \times 10^{10}$ & $ 1.4 \times 10^6$ \\
D5e11-A05-RT & $5 \times 10^{11}$ & 0.5 & $ 6.9\times 10^6$ & $2-3\times 10^{12}$ & $ 110$ & $0.1\times 10^9$ & $6.2 \times 10^{9}$ & $1.2\times 10^6$ \\
D5e11-A08-RT & $5 \times 10^{11}$ & 0.8 & $ 6.7\times 10^6$ & $1-2\times 10^{12}$ & $ 100$ & $3.5\times 10^8$ & $6.1 \times 10^{9}$ & $1.2\times 10^6$ \\
D1e11-A05-RT-PR & $10^{11}$ & 0.5 & $ 6.1\times 10^6$ & $6\times 10^{11}$ & $ 550$ & $3.9\times 10^9$ & $3.5 \times 10^{9}$ & $ 0.85 \times 10^5$ \\
D1e11-A08-RT-PR & $10^{11}$ & 0.8 & $ 6.0\times 10^6$ & $4.7\times 10^{11}$ & $ 500$ & $2.4\times 10^9$ & $1.78 \times 10^{9}$ & $ 0.78 \times 10^5$ \\
D5e11-A05-RT-PR & $5 \times 10^{11}$ & 0.5 & $ 6.7\times 10^6$ & $2-3\times 10^{12}$ & $100$ & $5.5\times 10^8$ & $0.69 \times 10^{9}$ & $0.6 \times 10^5$ \\
D5e11-A08-RT-PR & $5 \times 10^{11}$ & 0.8 & $ 6.9\times 10^6$ & $2-3\times 10^{12}$ & $70$ & $4.2\times 10^8$ & $0.46 \times 10^{9}$ & $0.52 \times 10^5$ \\
\hline 
\end{tabular}
\tablefoot{
Each model has same velocity of the accreting plasma before the shock, $v_{acc} = 500 \rm \; km/s$, and post-shock velocity, $v_{post}\sim 100-200 \rm \; km/s$.
}
\end{table*}

\subsection{Radiative heating of pre-shock plasma}
\label{sec.2.2}

\begin{figure}
  \resizebox{\hsize}{!}{\includegraphics{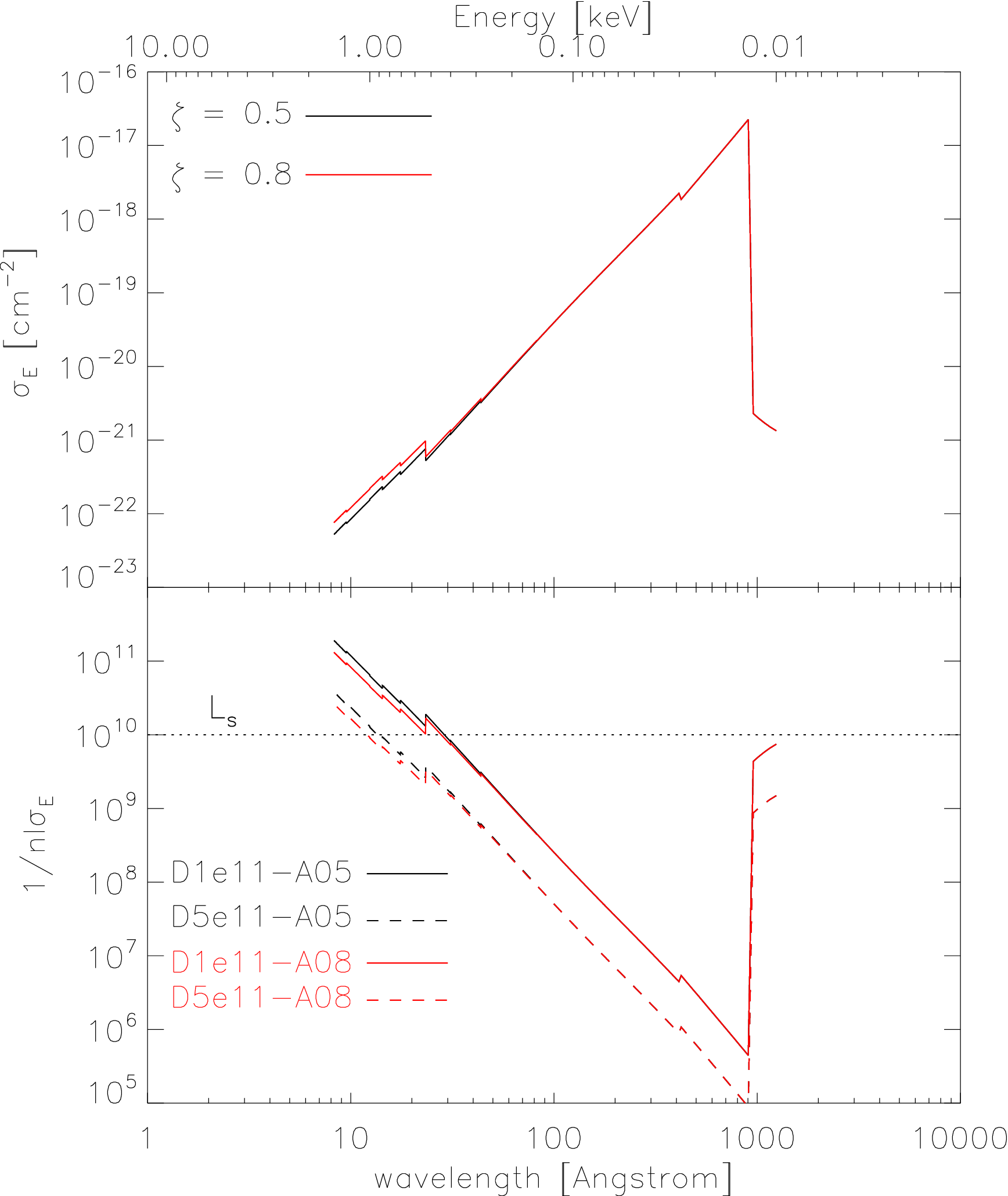}}
  \caption{Top panel: photonionization cross section as a function 
  of the wavelength
  from the PINTofALE library \citep{Kashyap2000BASI...28..475K},
  assuming the heavy element abundance of 0.5 (black line) and 0.8 (red line). Bottom panel: mean
  free path as a function of the energy, for the two density used in our models (solid line for $n_{acc}=10^{11}$ cm$^{-3}$, and 
  dashed lines for $n_{acc}=5 \times 10^{11}$ cm$^{-3}$). 
  The dotted horizontal line ($L_s$) indicates the
  characteristic length of the accretion stream equal to its section
  radius.} \label{fig:sig.Rad}
\end{figure}

The main aim of this work is to study the effect of irradiation of
the infalling material by the emission arising from the post-shock
plasma in the impact region. To this end, we developed a numerical
code (hereafter RT code) to derive the fraction of irradiating
energy absorbed by cold and dense plasma. Our analysis focuses in
the range of energy between the ionization edge of hydrogen ($\rm
13.6\;eV$) and the soft X-ray emission (up to $E \rm \sim 1.3\;keV$),
which is characteristic for this type of phenomenon.  We assumed
that the only source of radiative heating is the hot
post-shock plasma with $T> 1$ MK. We investigated the possible heating
only of the pre-shock material, due to this X-ray radiation.
Absorption of the cold pre-shock material is regulated by the plasma density
along the photons path and by the
photoionization cross section ($\sigma_E$) that is shown in Fig.
\ref{fig:sig.Rad} (Panel (a)).
Panel (a) shows that the probability 
for a high energy photon to be absorbed is smaller than that 
for a low energy photon. Panel
(b) shows that the mean free path (computed with fixed values of
 densities and metal abundance 
related to the different models) of a high-energy photon is greater
than the characteristic length of the accretion stream ($L_s
\sim 10^{10} \rm \; cm$), whereas the mean free path of a photon with
low energy can be smaller than $L_s$ even by orders of magnitude.
As a result, in general the accretion stream can be considered
to be optically thick to UV and soft X-ray radiation.

The RT code is written in C programming language and
is coupled to the PLUTO code. In particular the results of the HD
simulations from PLUTO are used as input for the RT
code to calculate the energy absorption (and the related heating
function $H(z)$; see Eq.\ref{eq:cons.ene}) along the accretion
stream. The absorption is solved in 1-D. The RT code consists
of the following steps:

\begin{enumerate}
\item \textbf{Reading the HD variables.} As a first step the code reads the
input, namely the profiles of temperature (T), density ($\rho$),
and velocity (v) produced by PLUTO, and the cross section ($\sigma$)
(Fig. \ref{fig:sig.Rad}) derived from the function ISMTAU
of the PINTofALE library \citep{Kashyap2000BASI...28..475K}.

\item \textbf{Spectral synthesis.} The code identifies the slab of
shocked hot plasma by selecting the plasma with $T>1$ MK. Then it 
synthesizes
the emitted spectrum (in the energy range between $\sim 0.01-1.3$ keV) from each grid point of this region
(assumed to be optically thin), using the CHIANTI atomic database
\citep{Dere1997A&AS..125..149D, Landi2013ApJ...763...86L}, and
integrates all the contributions to obtain the total emerging spectrum:

\begin{equation}
        \centering
        I_E(0)=\int_{post} G(T,E)\,n^2_e\,A\,dz,
        \label{eq:Int.Emess}
\end{equation}

\noindent
where $post$ is the length of the post-shock zone, $n_e$ is the
electron number density, $G(T,E)$ is the spectrum of an
isothermal optically thin plasma (i.e. the emissivity of
the plasma vs. its temperature, $T$, and the  photon
energy, $E$, per unit of emission measure), $A$ is the section
 area of the stream. Here
we assumed that one half of the total luminosity emitted by the hot
slab, $I_E(0)$, goes toward the star surface and is unobservable,
and the other half propagates outwards along the 1-D downflowing 
plasma
(plane parallel approximation, see Sec. \ref{sec.2.4}).

\item \textbf{Absorption.} The code calculates the optical
thickness ($\tau_E$) and computes the transmitted spectrum ($I_E(z)$)
for every grid point along the accreting pre-shock material with:

\begin{equation}
	\centering \tau_{E} = \int_l \sigma_E\,n_e(z)\,dz,
	\label{eq:spess.ott}
\end{equation}
\begin{equation}
        \centering
        I_E(z) = I_E(0) \,e^{-\tau_E(z)},
        \label{eq:abs.spect}
\end{equation}

\noindent
where $\sigma_E$ is the cross section as a function of the energy
(Fig. \ref{fig:sig.Rad}), $n_e(z)$ is the number density as a
function of the coordinate $z$, $l$ is the distance from the shocked plasma
region, $dz$ is the length element, $I_E(0)$ is the spectrum
vs. energy from the post-shock zone. 

\item \textbf{Absorbed energy and volumetric heating.} 
Finally the code
calculates the absorbed energy along the stream and, assuming
that all this energy produces local heating, 
it calculates the volumetric heating $H_t(z)$ as

\begin{equation}
        \centering
        \frac{dI_E}{dz}= - n_e\,\sigma_E\,I_E(z),
        \label{eq:ener.assor}
\end{equation}

\begin{equation}
        \centering
        H_t(z) = -\int_E \frac{1}{A}\frac{dI_E}{dz}\;dE= \int_E\frac{n\,\sigma_E}{A}\,I_E(z)\;dE,
        \label{eq:risc.vol}
\end{equation}

\noindent
the $H_t(z)$ term is in units of $\rm erg \; s^{-1}\; cm^{-3}$.

\end{enumerate}

As mentioned before, in our study we assumed $\beta \ll 1$, and our
simulations describe the evolution of a single fibril. Thus
the whole accretion stream can be considered as formed by
many of these elementary fibrils. Each of them is independent
from the others and, in principle, may have a different
evolution phase, even though they have the same type of evolution
\citep{Orlando2010}.

In order to calculate the average volumetric heating rate, $H(z)$, of a
fibril in the central part of the stream, we considered
the emission from the surrounding fibrils (each with its
own phase). Then we calculated the absorption and the
volumetric heating rate along the fibril. This (central) fibril
is expected to get the combined irradiation due to all the fibrils
around; since each of them is in a different phase of
evolution, we consider the sum of the emission at all the
times (or the time average) of one fibril, equal to the average
emission of the whole surrounding fibrils in a time lapse (or space
average). In Sect. \ref{sec.2.4} we discuss in detail the
limits of this assumption.

\subsection{Coupling the radiative heating and hydrodynamic computations}
\label{sec.2.3}

The HD and absorption feedback computations are coupled through the 
volumetric heating
function, $H(z)$ (see Eq. \ref{eq:cons.ene}).
To obtain a self-consistent solution of the coupled equations, we
adopted the following iterative method:

\begin{enumerate}

\item First the HD Eqs. \ref{eq:cons.mass}
- \ref{eq:cons.ene} are solved without the heating term $H(z)$ in the energy
conservation equation using the PLUTO code (see Sec. \ref{sec.2.1});

\item The resulting profiles of temperature, pressure,
density, and velocity, are used as input for the RT code (see Sect.
\ref{sec.2.2}) to solve the absorption along the accretion stream, and to
calculate the time-average $H(z)$ function;

\item A new, otherwise identical, simulation is performed using the
same parameters of the previous one, but now including the
$H(z)$ function calculated in the step 2; in this simulation
therefore the feedback of absorption of UV and soft X-ray emission
arising from the shock heated plasma is included in the dynamics 
and energetics of the stream;

\item Since steps 1, 2, and 3 are not performed in a
physically self-consistent model, the second and third
steps are iterated until the convergence of the HD solution is obtained;
the convergence is checked by analyzing the time
average of the emission measure distribution vs. temperature (see
Sec. \ref{sec.3.3}) of the plasma, and the volumetric heating
function $H(z)$. The iterative method is stopped when the relative
difference of these quantities are below $1 \%$ between two consecutive 
iterative steps.

\end{enumerate}

\subsection{Limits of the model}
\label{sec.2.4}

Here we discuss the assumptions and hypothesis made in this work.

\begin{itemize}

\item \textbf{Radiative losses from optically thin plasma}. As
mentioned in Sect. 2.1, our model treats approximately the cooling 
of the pre-shock material (which, in
general, is neither completely thin nor thick). In the models
D1e11-A05-RT, D1e11-A08-RT, D5e11-A05-RT, and D5e11-A08-RT (see 
Table~\ref{tab:par.mod}) the radiative
cooling of the infalling plasma is neglected, thus maximizing
the effect of radiative heating.  Conversely, in the models
D1e11-A05-RT-PR, D1e11-A08-RT-PR, D5e11-A05-RT-PR and D5e11-A08-RT-PR 
the radiative cooling is overestimated
by the assumption of optically thin plasma, thus minimizing the
effect of heating.  To further investigate this
issue, we calculated the absorption of a spectrum between $\rm 0.01
- 1.3\; keV$ emitted by an isothermal plasma, for different values
of temperature. Fig. \ref{fig:Abs.perc} shows the characteristic
thickness at which $\approx 90\%$ of the luminosity is absorbed
assuming a stream density $n_{str} =10^{11}\rm \; cm^{-3}$;
As we will
show in Sect. \ref{sec.3}, the dimension of the pre-shock material,
that is significantly heated by irradiation of X-rays from
the post-shock plasma, is indeed comparable with these values of $L_{90\%}$.
This evidence confirms that the pre-shock material
cannot be considered to be completely thick or thin. Thus
we expect that the temperature and density of the radiatively heated pre-shock
material are in between the two cases considered in our study.
Nevertheless, our approach allowed us to bracket the effect of
radiative heating of the unshocked stream and the corresponding
contribution to emission in UV and X-ray bands.

\begin{figure}
  \resizebox{\hsize}{!}{\includegraphics{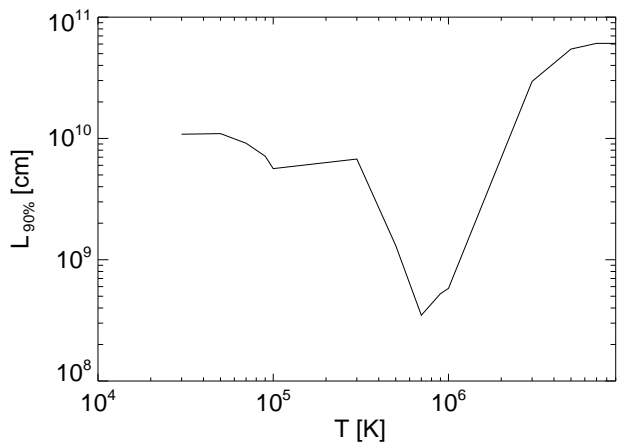}}
  \caption{Characteristic thickness at which about $90\%$ of the radiation intensity is absorbed along the pre-shock plasma, with 
$n_{acc} =10^{11}\rm \; cm^{-3}$.}
  \label{fig:Abs.perc}
\end{figure}

\item \textbf{Plasma heating capacity}.
The ionization state of the plasma changes its ability to transform
the absorbed radiation into heat. In our model we assumed that the
plasma converts all the absorbed radiation in heating, but the
degree of ionization of the plasma at the disk inner edge, and then
that of the infalling plasma, is still an issue largely
debated in the literature. We can suppose that the plasma in the
disk may be ionized for a small fraction, and during the fall this
fraction grows. So, due to this fact, the photoionization cross
section of the material may change as the ionization change.
On the other hand, we used a fixed cross section of a non-ionized 
gas with metal abundance of 0.5 of the solar value.
From studies of \citet{Furlanetto2010MNRAS.404.1869F} we estimated
that, for a plasma with an ionization fraction of $10\%$ and for
energy of the ionization photon greater than $E\sim 1 \;$ keV, the
fraction of energy that is converted into heating is about $70\%$.
For lower energies of the ionization photon this fraction increases.
These considerations assure that, at the first order, our results
provide an appropriate description of the system.

\item \textbf{Geometry of the system}.
In our models we assumed that the accretion flow extends radially
up to about $0.7\; R_*$ ($R_* = 9.048 \times 10^{10} \;$ cm).
However the plasma streams follow the magnetic field lines which,
at large distances from the star, have a configuration close to
that of a dipole. A magnetic arc that connects the star with the
disk can have a length of about $(3 - 4)\; R_* \sim (3 - 4)
\times 10^{11} \;$ cm. On the other hand, as discussed in
Sect.~\ref{sec.3}, the radiative heated precursor can be smaller
than the arc extension by more than an order of magnitude (see Table
\ref{tab:par.mod}).  Thus our assumption to neglect the curvature
of the accretion stream does not affect significantly the description
of the precursor. As discussed in Sec. \ref{sec.2.2}, we assumed the 
plane parallel approximation;
this approximation is valid only in the inner part of the stream
(namely far away from the stream border) and for distances from the
stellar surface much smaller than the cross-section radius of the
stream.  By comparing the characteristic thickness of the precursor
(see Sect. \ref{sec.3}) with the adopted cross-section radius of
the stream ($r_f \sim 3.4 \times 10^{10} \;$ cm), we
conclude that our assumption does not affect significantly
the results.

\end{itemize}

\section{Results}
\label{sec.3}

\subsection{The reference model}
\label{sec.3.1}

As a first step, we performed a simulation of an accretion stream
impact onto a stellar surface neglecting the radiative heating.
An estimate of the expected slab parameters can be made
using the strong shock limit \citep{zeldovich1967}. We report the
equations below for convenience of the reader:

\begin{equation}
	\centering
	v_{post}=\frac{v_{acc}}{4} \qquad  n_{post}=4n_{acc},
	\label{eq:vel.den}
\end{equation}
\begin{equation}
	\centering
	T_{post}=\frac{3}{32}\frac{\mu m_H}{k_B}v^2_{acc}\approx 1.4 \times 10^{-9}v^2_{acc},
	\label{eq:temp.asp}
\end{equation}
\begin{equation}
	\centering	
	\tau_{cool}=\frac{1}{1-\gamma}\frac{P}{n_en_H\Lambda(T)} \sim 2.5 \times 10^3 \frac{1}{\zeta} \frac{T^{3/2}_{post}}{n_{post}},
	\label{eq:raf.asp}
\end{equation}
\begin{equation}
	\centering	
	\tau_{cross} = \frac{L_{post}}{v_{post}} \equiv \tau_{cool},
\end{equation}
\begin{equation}
	L_{post}\equiv \tau_{cool} v_{post}  = 1.7\times 10^{-11} \frac{1}{\zeta} \frac{v_{acc}^{4}}{n_{acc}},
	\label{eq:spes.asp}
\end{equation}

\noindent
where $v$ is the velocity, $n$ is the number density, $T$ is the temperature,
$P$ is the thermal pressure, $\tau_{cool}$ is the cooling time of
the slab, $\tau_{cross}$ is the crossing time of the slab, $L_{post}$
is the maximum thickness of the slab. Table \ref{tab:par.asp}
reports the expected values for the six models explored,
calculated according to the Eq. \ref{eq:vel.den} - \ref{eq:spes.asp}.

\begin{figure}
  \resizebox{\hsize}{!}{\includegraphics{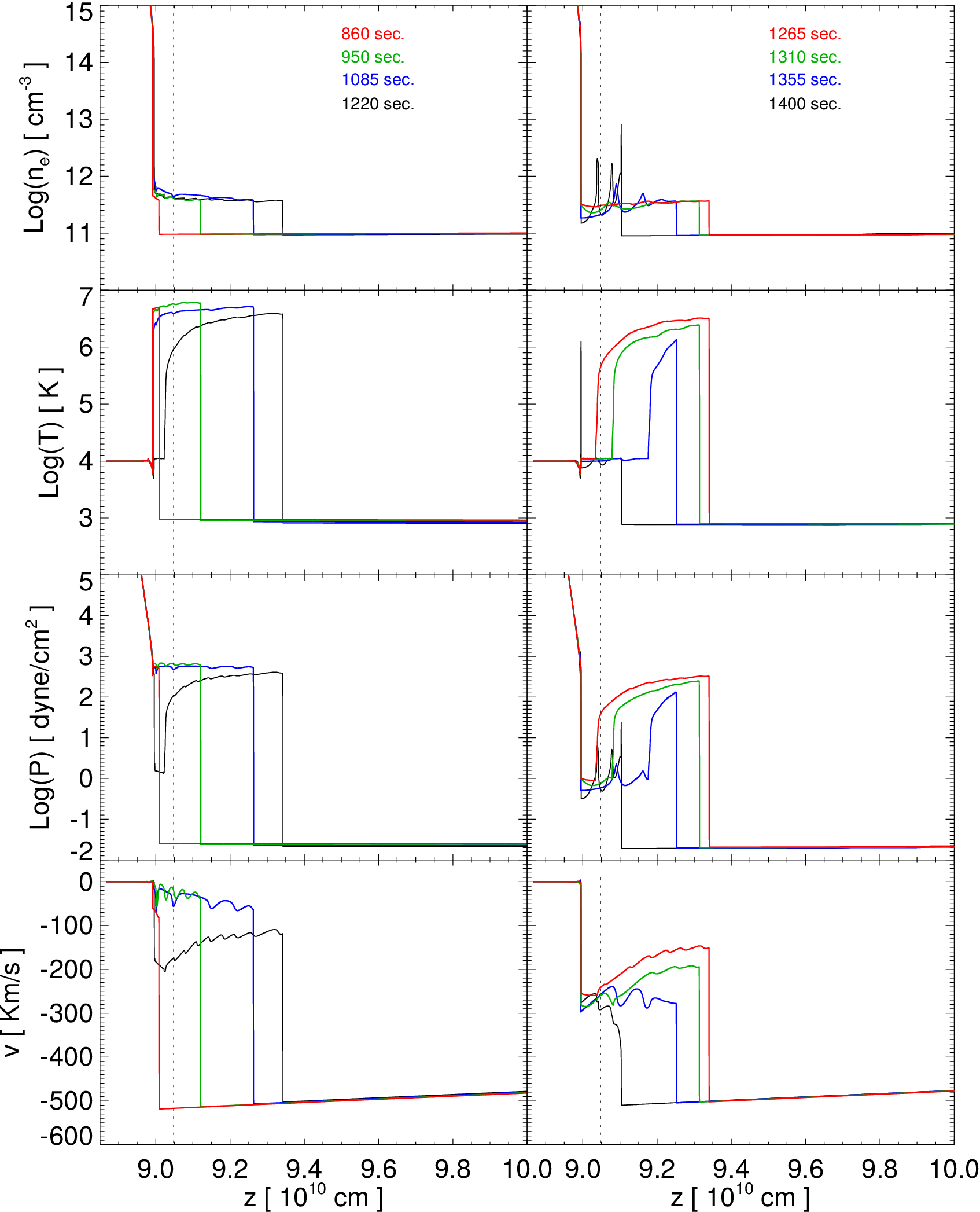}}
  \caption{Time evolution of plasma density, temperature, pressure,
  velocity of the post-shock plasma in the case of the model 
  D1e11-A05.
  The panels on the left show the expansion phase of the slab, while
  the panels on the right show the collapse phase. The vertical
  dashed lines mark the position of the unperturbed transition
  region.}
  \label{fig:mod.w/o_rt}
\end{figure}

\begin{table*}
\caption{Expected parameters of the post-shock zone for the models with different stream density.} 
\label{tab:par.asp} 
\centering
\begin{tabular}{ccccc}
\hline\hline 
Expected Values & D1e11-A05 & D1e11-A08 & D5e11-A05 & D5e11-A08 \\ 
\hline 
$v_{post}\;$ [km/s] & 125 & 125 & 125 & 125 \\ 
$T_{post}\; $[K] & $3.5\times 10^6$ & $3.5\times 10^6$ & $3.5\times 10^6$ & $3.5\times 10^6$ \\
$n_{post}\; $ [cm$^{-3}$] & $4\times 10^{11}$ & $4\times 10^{11}$ & $2 \times 10^{12}$ & $2 \times 10^{12}$ \\
$\tau_{cool}\;$ [s] & 327 & 205 & 66 & 41 \\ 
$L_{post}$ [cm] & $2.1\times 10^9$ & $1.3\times 10^9$ & $4.3\times 10^8$ & $2.6\times 10^8$\\ 
\hline 
\end{tabular} 
\end{table*}

After the impact of the accretion stream onto the stellar
surface, a hot (few million degrees) and dense ($n_{post}\sim 4 \times
10^{11}\;$ cm$^{-3}$) slab of shock-heated plasma forms at the base
of the stream. The slab is rooted in the chromosphere where the ram
pressure of the stream, $P_{ram} = \rho \;v^2$, equals the thermal
pressure, $P_T$, of the chromosphere \citep{Sacco2008}. As discussed
by several authors \citep[e.g.][]{Sacco2008}, the slab is characterized
by quasi-periodic oscillations of the shock-front. Fig.
\ref{fig:mod.w/o_rt} shows the evolution of density, temperature,
pressure, and velocity during one of these oscillations for the
model D1e11-A05 (see Table~\ref{tab:par.mod}). The left panels show
the so-called ``expansion phase'' during which the slab forms. The
continuous downfalling plasma powers the gradual growth of the slab.
This process continues until the radiative losses at the base of
the slab trigger thermal instabilities there that cause a rapid
decrease of temperature and pressure in some parts of the
slab which appear as cold structures between portions
of hot plasma. This is the so-called ``collapse phase'' in which
the intense radiative cooling robs the slab of pressure support,
causing the material above the cooled layers to collapse back before
the slab expands again (the relevant evolution is showed on the
right panels of Fig.  \ref{fig:mod.w/o_rt}). The evolution is then
characterized by alternating phases of expansion and collapse of
the post-shock region. In the case of model D1e11-A05, the slab has
a quasi-periodic evolution with a characteristic time of $P_o\sim
550 \;$ s, it reaches a maximum thickness of $L_{post} = 3.9 \times
10^9 \;$ cm , and a maximum temperature of $T_{post} = 6.6 \times
10^6 \;$ K.

The other models (D1e11-A08, D5e11-A05, D5e11-A08; see
Table~\ref{tab:par.mod}) show an analogous evolution of the
post-shock region but, because of the different density and
metallicity, the slab has different characteristics with respect to
model D1e11-A05 (see Table \ref{tab:par.mod}). As showed
in Eq. \ref{eq:vel.den} - \ref{eq:spes.asp} the final velocity and
temperature depend only on the initial velocity; the cooling time
and the slab thickness depend on the whole set of parameters
(namely $v_{acc}$, $n_{acc}$, and $\zeta$); so, higher
values of density and/or metallicity imply stronger radiative losses
and, therefore, smaller cooling time and slab thickness; furthermore,
in case of higher density, because of its higher ram pressure,
$P_{ram}$, the post-shock zone results rooted deeper in the
chromosphere.  We found a good agreement between the values derived
from the models (see Table \ref{tab:par.mod}) and those expected
(see Table \ref{tab:par.asp}).

\subsection{Effects of the radiative heating feedback}
\label{sec.3.2}

\begin{figure}
  \resizebox{\hsize}{!}{\includegraphics{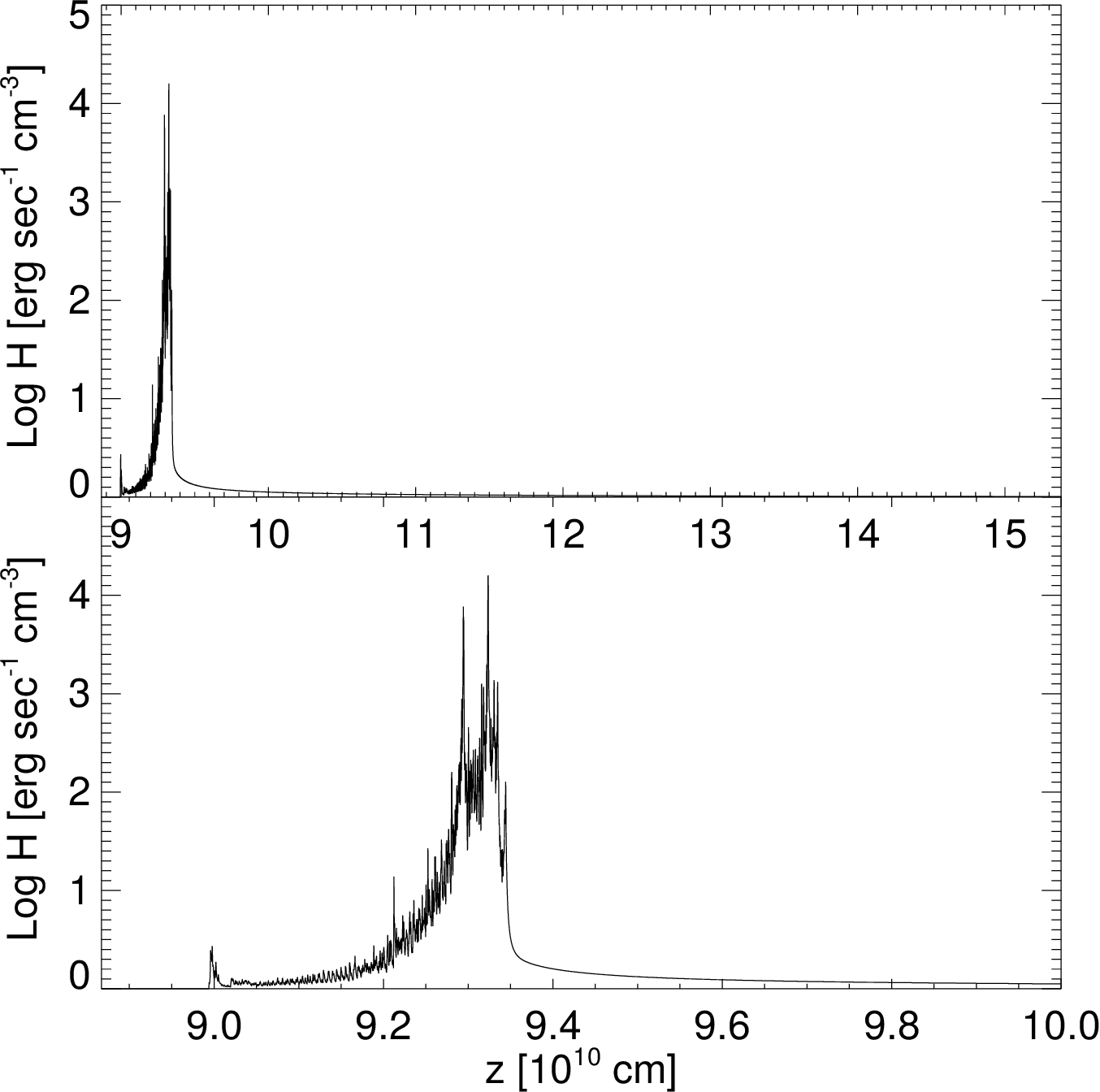}}
  \caption{Volumetric heating rate of model D1e11-A05 as a function of 
  the position, $z$. Top panel: the function in the whole
  domain; Bottom panel: a zoom of the function in the shock region. }
  \label{fig:risc.vol}
\end{figure}

Fig. \ref{fig:risc.vol} shows the radiative heating function
$H(z)$ derived with the RT code for the model D1e11-A05; the function
shows a peak which corresponds to the maximum extension of the slab.
The contact discontinuity between the chromosphere and
the slab is at $z = 9\times10^{10} \;$ cm. After the peak, the
$H(z)$ function shows a typical exponential trend due to the
absorption (see Eq. \ref{eq:abs.spect}).

\begin{figure}
  \resizebox{\hsize}{!}{\includegraphics{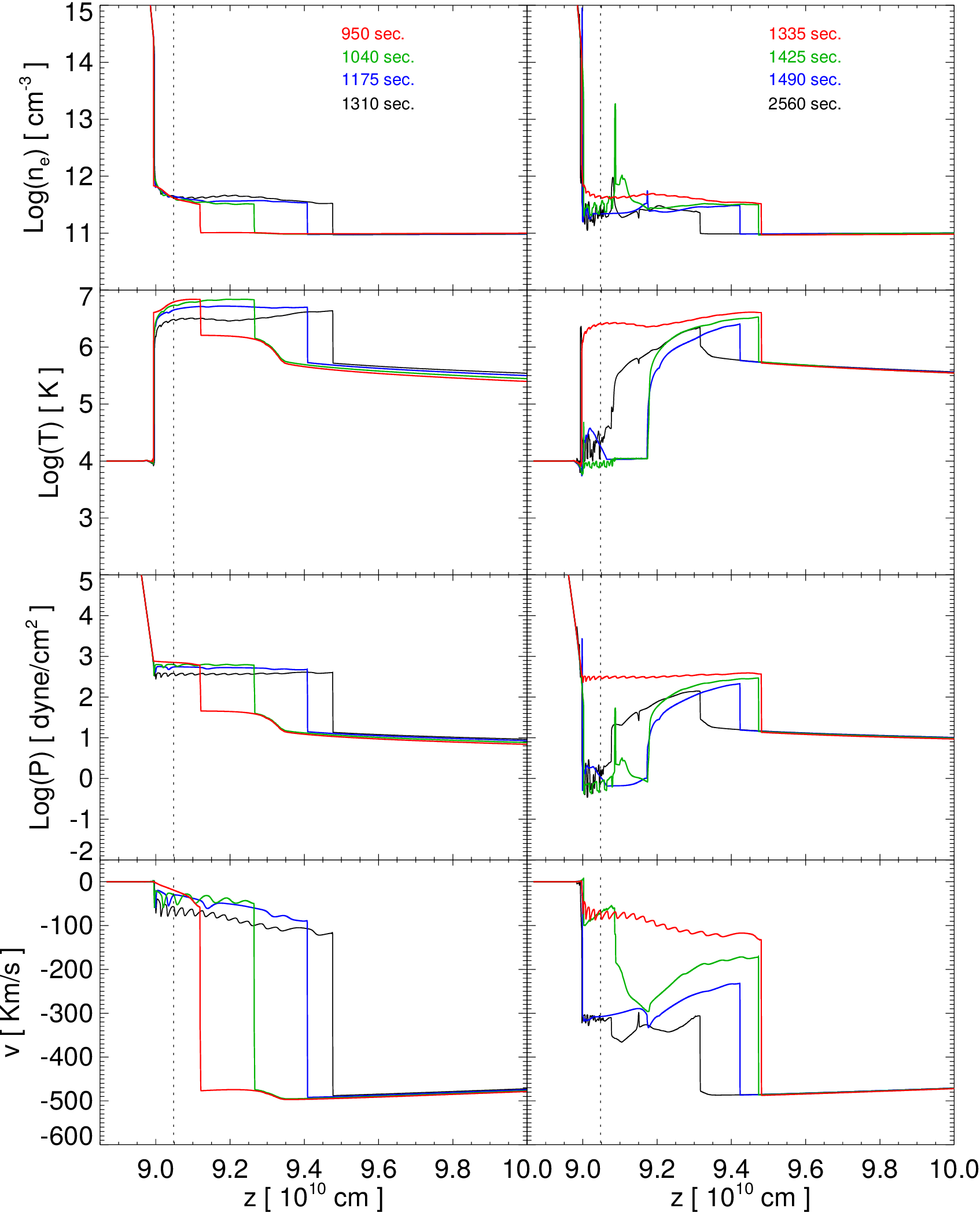}}
  \caption{As in Fig.~\ref{fig:mod.w/o_rt} for model D1e11-A05-RT
  including the heating function, $H(z)$, due to the X-ray absorption
  of the accretion stream and neglecting the radiative cooling in
  the unshocked stream.} \label{fig:mod.w_rt}
\end{figure}

\begin{figure}
  \resizebox{\hsize}{!}{\includegraphics{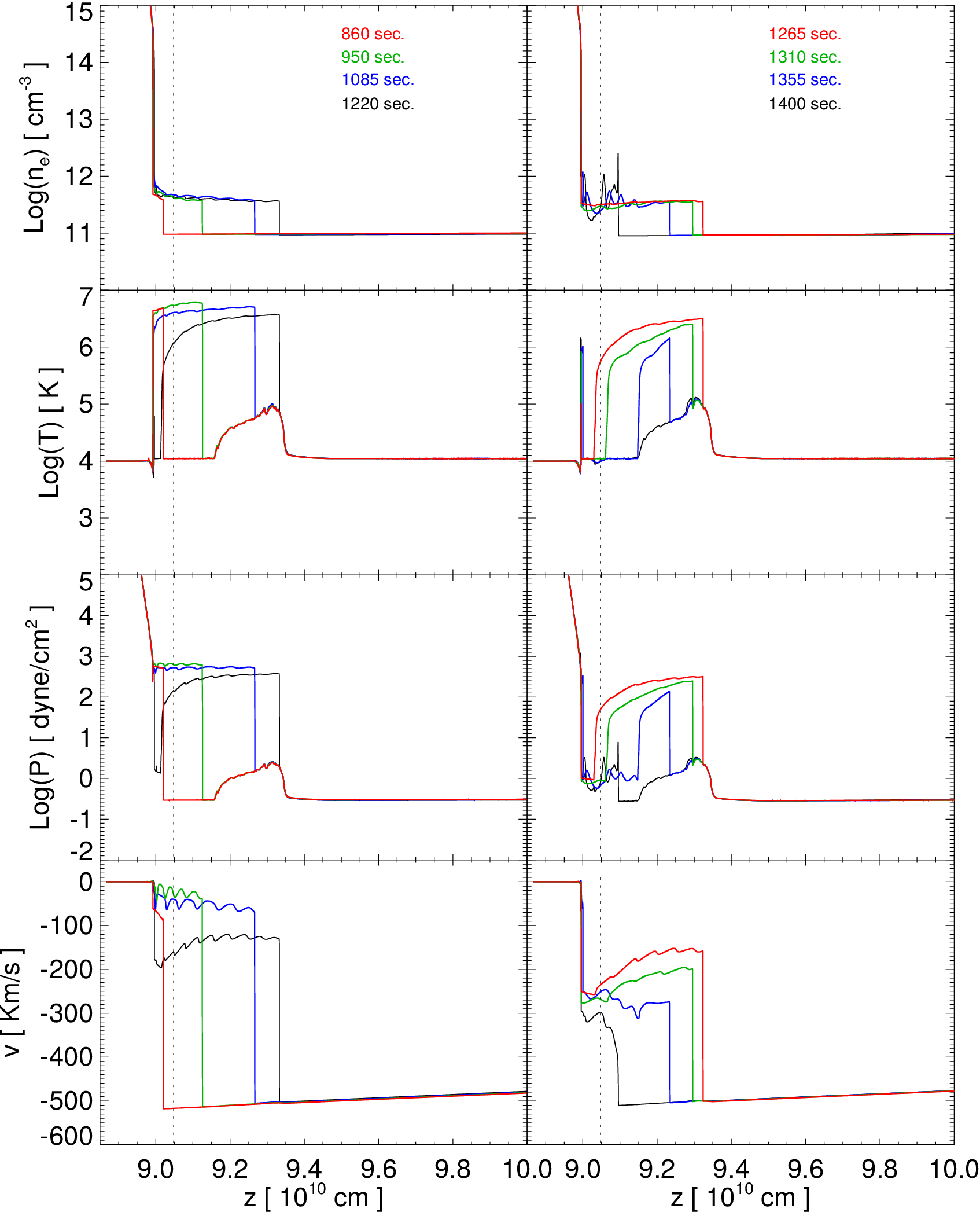}}
  \caption{As in Fig.~\ref{fig:mod.w/o_rt} for model D1e11-A05-RT-PR
  including both the radiative heating and the radiative losses
  from optically thin plasma in the unshocked stream.} \label{fig:mod.w_rt_pr}
\end{figure}

As explained in Sec. \ref{sec.2.3}, we used the volumetric
heating rate, $H(z)$, derived above to perform new HD simulations
but now including the heating of the unshocked stream due to
absorption of UV and X-ray emission arising from the post-shock
plasma. For each models (D1e11-A05, D1e11-A08, D5e11-A05 and D5e11-A08),
we explored the two extreme limiting cases discussed in
Sect.~\ref{sec.2.1}: in the first we neglected any radiative cooling
of the material of the unshocked stream, in the second we assumed
radiative losses from optically thin plasma in the unshocked stream.
Fig. \ref{fig:mod.w_rt}, in analogy with Fig. \ref{fig:mod.w/o_rt},
shows the evolution of density, temperature, pressure, and velocity
for a stream with initial density $n = 10^{11}$~cm$^{-3}$
and for the case neglecting the radiative cooling in the unshocked
stream (run D1e11-A05-RT). We found that the evolution
of the post-shock zone is roughly the same as that found in run
D1e11-A05 which neglects any radiative heating effect (compare Figs. 
\ref{fig:mod.w/o_rt} and \ref{fig:mod.w_rt}). This was expected because, in the unshocked
stream, the ram pressure is much larger than the thermal pressure,
$P_{ram} \gg P_T$, so that the heating of the unshocked material 
has a negligible effect on the dynamics of the post-shock region.
In Fig. \ref{fig:mod.w_rt_pr} we show the evolution of density,
temperature, pressure and velocity if we assume that the
unshocked material of the stream is optically thin (run D1e11-A05-RT-PR). Also in this case we found that the evolution of
the hot slab is roughly the same as in runs D1e11-A05 and D1e11-A05-RT.

In runs D1e11-A05-RT and D1e11-A05-RT-PR the slab reaches about the same
maximum temperature as D1e11-05, it has a maximum thickness of
$L_{post}\sim 4 \times 10^9 \;$ cm and the characteristic repetition
time is $P_o\sim 550 - 600\;$ s. The three models differ mainly in the
development of a hot region immediately before the accretion shock,
a so-called precursor. In run D1e11-A05-RT (namely that neglecting
any radiative losses in the pre-shock stream) the downfalling
plasma is gradually heated as it approaches the stellar surface:
its temperature increases from $T_{acc}\approx 10^3$~K (at the upper
boundary) up to $T_{prec} \sim 1.4 \times 10^6\;$ K immediately
before the accretion shock. The precursor is characterized
by a bump in temperature at $9.1 \times 10^{10} < z < 9.3\times
10^{10}$~cm due to the peak in the volumetric heating at $z \approx
9.25\times 10^{10}$~cm (see Fig. \ref{fig:risc.vol}). The portion
of precursor with $T > 10^5\;$ K extends up to $L_{prec} \sim 2.5
\times 10^{10}\;$ cm above the stellar surface.

In run D1e11-A05-RT-PR (namely that assuming optically thin plasma
in the unshocked stream), the heating of the downfalling plasma is
balanced by the radiative losses for temperatures above $10^4$~K.
As a result, the precursor is represented by a temperature
bump between $z \approx 9.1\times 10^{10}$~cm and $z \approx 9.3\times
10^{10}$~cm, corresponding to the position of the peak in the
volumetric heating (see Fig. \ref{fig:risc.vol}). The precursor
has a maximum temperature $T_{prec}\sim 0.9 \times 10^5\;$ K lower
than that in run D1e11-A05-RT and extends with $ T> 10^4 \rm \;K$
up to $L_{prec}\sim 3.5 \times 10^{9}\;$ cm above the stellar
surface. In this latter case, the thickness of the precursor depends
on the ratio between the heating due to the irradiation and the
radiative losses in the unshocked plasma: the stronger the heating
the larger the precursor. Given its temperature and density, in
both cases examined, the precursor is expected to be a strong source
of UV emission.

\begin{figure}
  \resizebox{\hsize}{!}{\includegraphics{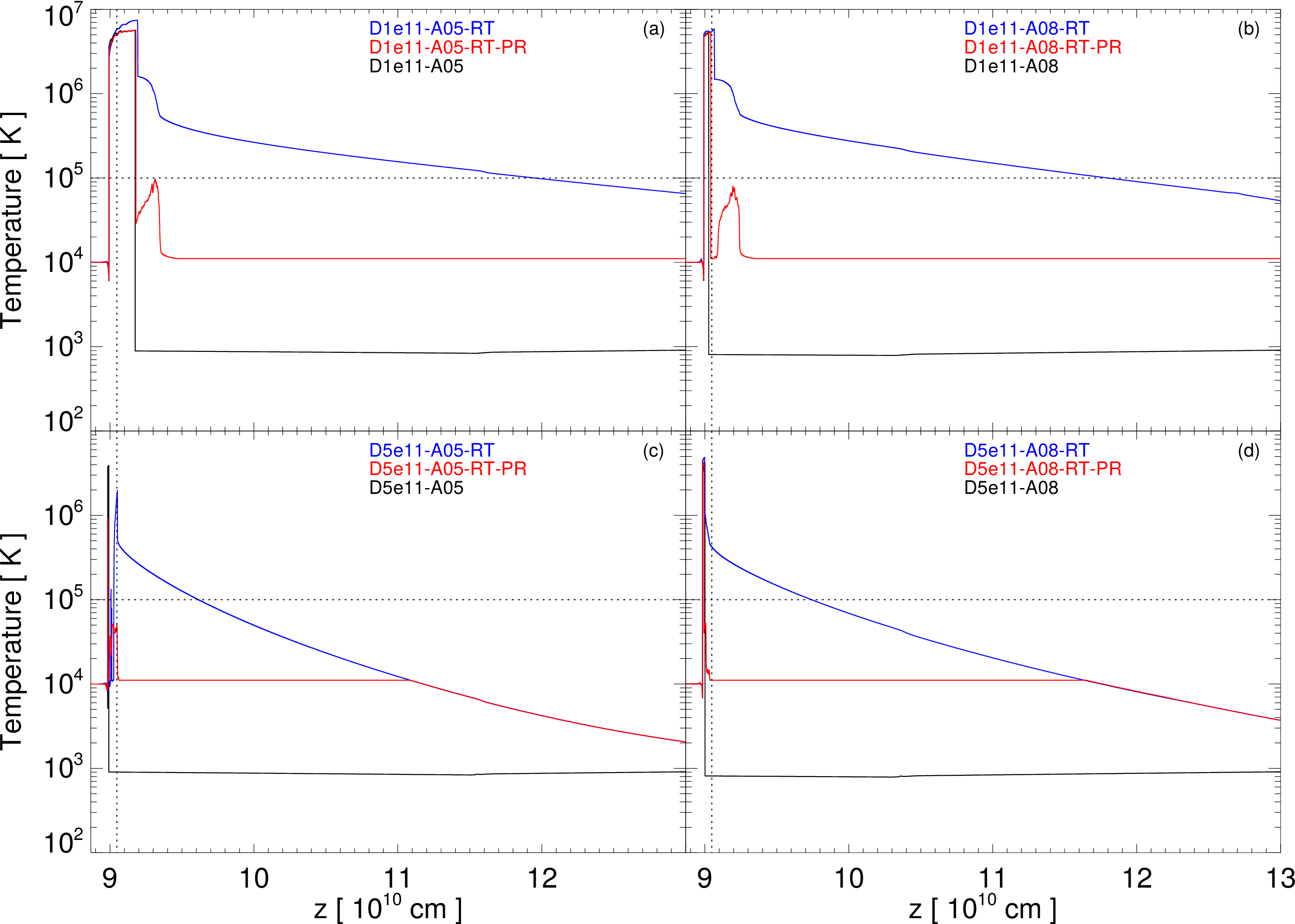}}
  \caption{Temperature profiles for models with stream density
  $n_{acc} = 10^{11}$ cm$^{-3}$ (upper panels) and $n_{acc} = 5\times
  10^{11}$ cm$^{-3}$ (lower panels). The vertical dashed lines mark
  the position of the stellar transition region, the horizontal
  dashed line indicates a temperature of $10^5\;$ K.} \label{fig:prec.tot}
\end{figure}

We found similar results in the case of models\footnote{Runs D1e11-A08, D5e11-A05, D5e11-A08, D1e11-A08-RT,
D5e11-A05-RT, D5e11-A08-RT, D1e11-A08-RT-PR, D5e11-A05-RT-PR and
D5e11-A08-RT-PR.} with an accretion stream with metal
abundance $\zeta = 0.8$ and/or initial density $n = 5\times
10^{11}$~cm$^{-3}$: the radiative heating causes the development of
a precursor in the region immediately before the accretion shock.
Fig. \ref{fig:prec.tot} shows the temperature profiles in the whole
domain, at the same time, for the twelve models considered.
When the radiative heating is taken into account in the computations,
each temperature profile shows a precursor which reaches the maximum
temperature near the post-shock region. The temperature profiles
decrease with the distance from the stellar surface, following the
trend of the volumetric heating rate function (see Fig.
\ref{fig:risc.vol}). Fig. \ref{fig:prec.tot} shows that either
denser streams or streams with higher metal abundance have a thinner
precursor.  Table \ref{tab:par.mod} reports the precursor parameters
of the simulations, where $L_{prec}$ is the characteristic size
of the precursor and $T_{prec}$ is the maximum plasma temperature
in the precursor zone.

\subsection{Distribution of emission measure versus temperature}
\label{sec.3.3}

\begin{figure*}
  \centering
  \includegraphics[width=17cm]{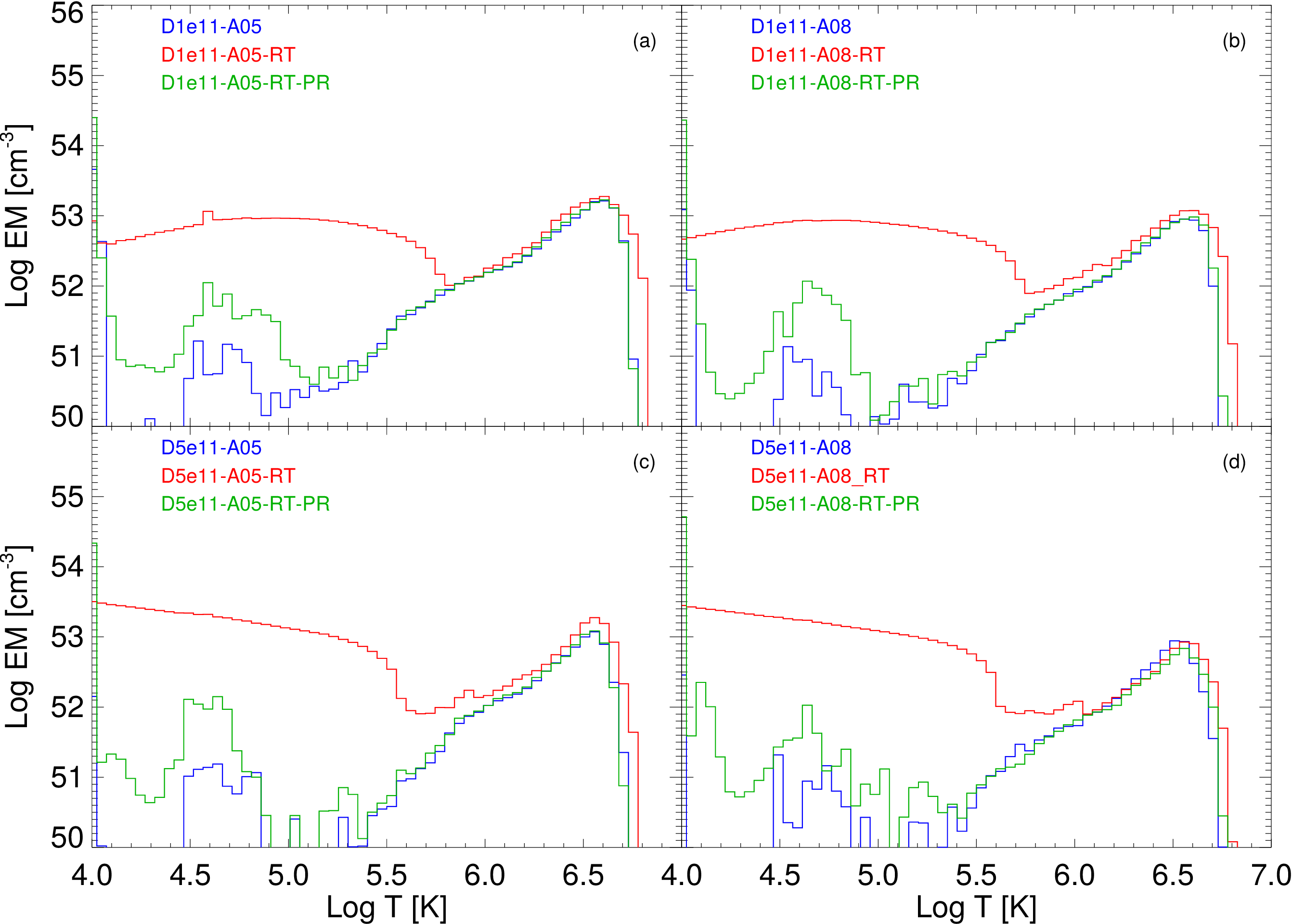}
  \caption{Emission measure distributions versus temperature 
  for models with stream density $n_{acc} = 10^{11}$ cm$^{-3}$
  (upper panels) and $n_{acc} = 5\times 10^{11}$ cm$^{-3}$ (lower
  panels). The figure shows the EM(T) derived for models either
  with (red and green lines) or without (blue) radiative heating
  and models either including (blue and green) or neglecting (red)
  the radiative losses from optically thin plasma in the unshocked
  stream.} \label{fig:EM.tot}
\end{figure*}

The distribution of emission measure versus temperature, $EM(T)
=\int_V n_e^2\, dV = \int_z n_e^2\,A\,dz$, is a useful tool that
gives us information about the components of plasma emitting at
various temperatures. We consider here the $EM(T)$
distribution of an accretion stream composed by several elementary
fibrils in random phases of evolution. Thus the resulting $EM(T)$ is obtained as the sum of all the contributions from the
fibrils and assuming a total accretion rate of $Log\; \dot{M}
= - 9.2 $ (with $\dot{M}$ in $\rm M_\odot/yr$). We do not expect, therefore, significant
variability of the $EM(T)$ distribution due to the alternating
phases of expansion and collapse of the post-shock region. The
accretion rate value considered is consistent with those inferred
from optical observations of three CTTSs, namely MP Mus, TW Hya and
V4046 Sgr; data from \citet{Curran2011A&A...526A.104C}).  The modeled
$EM(T)$ can be also directly compared with the $EM(T)$ distributions
derived from observations \citep[see e.g.][]{Argiroffi2009A&A...507..939A}.
For each model, the cross-section area A is derived from the adopted
accretion rate and from the stream density.  In particular, in the
case of streams with initial density $n=10^{11}$~cm$^{-3}$, we
adopted an area $A_{D1e11} = 3.7 \times 10^{21} \rm \;cm^2$
(corresponding to a filling factor of 3.6\%), whereas in
the case of streams with $n=5\times 10^{11}$~cm$^{-3}$, we adopted
$A_{D5e11} = 7.4\times 10^{20} \rm \;cm^2$ (filling factor
of 0.7\%).

Fig. \ref{fig:EM.tot} shows that for temperatures above one million
degree, all the distributions have roughly the same trend, in
particular they show a strong peak of EM at a temperature
of a few million degrees. This peak corresponds to the shock heated
plasma and is analogous to that studied by \citet{Sacco2010}.
For temperatures below one million degrees, the models show
different EM(T) distributions. In particular, models including the
radiative heating of the unshocked stream show values of EM at
temperatures around $T=10^5 \rm \;K$ higher than those in models
without radiative heating. These higher values of EM are due to the
precursor. On the other hand, Fig. \ref{fig:EM.tot} also shows that
models neglecting the radiative losses in the unshocked stream
predict a much higher and larger peak of EM at about $T=10^5 \rm \;K$.
As discussed in Sect.~\ref{sec.2.1}, models either with or without
the radiative losses in the unshocked stream represents the two
extreme limiting cases. We expect that more accurate distributions
of EM(T) calculated from models including self-consistently the
radiative transfer are in between the two cases considered here.
Since the emission measure depends on the density, the $EM(T)$ distributions
of models with high values of density (panel (c) and (d) of Fig.
\ref{fig:EM.tot}) show values of EM for $T < 1$~MK larger 
than models with lower density (see panel (a) and (b)).

\subsection{X-ray and UV luminosity of selected lines}
\label{sec.3.4}

\begin{figure*}
  \centering
  \includegraphics[width=17cm]{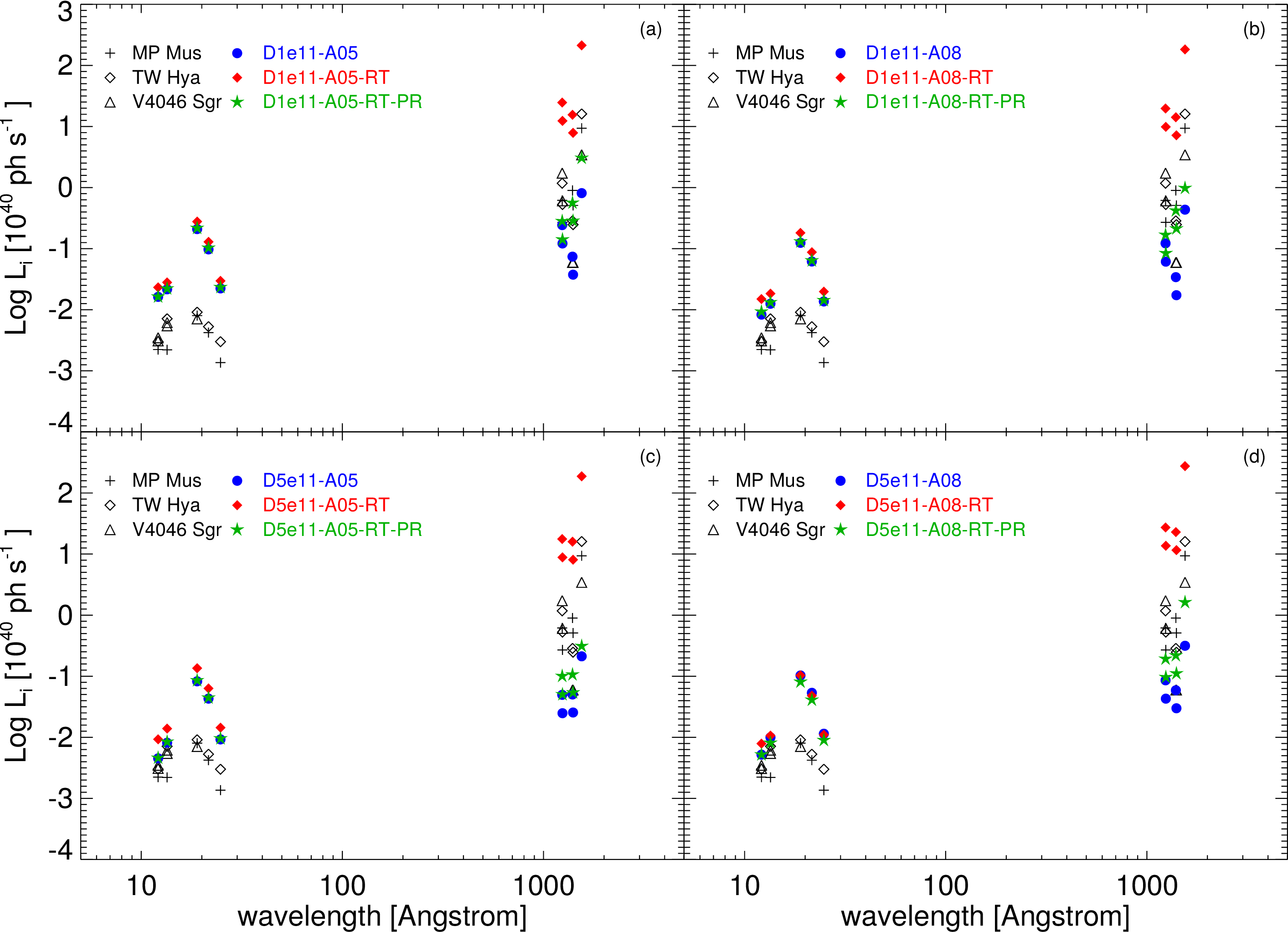}
  \caption{Luminosity of selected lines for three T-Tauri Stars (in 
  black), and luminosity synthesized from models with stream density
  $n_{acc} = 10^{11}$ cm$^{-3}$ (upper panels) and $n_{acc} = 5\times
  10^{11}$ cm$^{-3}$ (lower panels). The figure shows the luminosity
  from models either with (red and green lines) or without (blue)
  radiative heating and models either including (blue and green)
  or neglecting (red) the radiative losses from optically thin
  plasma in the unshocked stream.  Note that some of the symbols
  superimpose each other.}
  \label{fig:Lum.1}
\end{figure*}

We compared our results with UV and X-ray observations
\citep{Argiroffi2007A&A...465L...5A, Brickhouse2010ApJ...710.1835B,
Argiroffi2012,Ardila2013ApJS..207....1A} by synthesizing the
luminosity of selected lines (that are known to be strongly correlated
to accretion phenomena) from the emission measure distributions
derived in Sect.~\ref{sec.3.3}.  The lines considered are
listed in Table \ref{tab:Lum.rig.oss} - \ref{tab:Lum.rig.mod} -
\ref{tab:Lum.rig.mod2}.  In the X-ray band, we selected strong
emission lines usually ascribed to the post-shock region. In the
UV band, we selected lines that are known to have a correlation
with accretion phenomena \citep{Ardila2013ApJS..207....1A}.

In the following, we assumed that the line of sight is perpendicular
to the accretion stream. Under such assumption, the X-ray emission
originating from the slab might be partially absorbed by the stellar
chromosphere (because the slab is partially rooted in the chromosphere)
but not by the stream itself. On the other hand, the UV line emission
is not expected to suffer any significant absorption due to the
position of the precursor, it is over the stellar chromosphere so
the arising UV emission do not suffer any absorption.

Fig. \ref{fig:Lum.1} shows the luminosity of the selected lines in
UV and X-ray bands for the different models. The synthetic
luminosities in the figure have error bars equal to or below $10\%$
of the corresponding values. As for the $EM(T)$ distribution, we
do not expect any significant variability in the luminosity because
the stream is assumed to consist of many elementary fibrils, each
in a different random phase of evolution. This is also in agreement with
observations (see \citealt{Drake2009ApJ...703.1224D}). Our models without
radiative heating in the unshocked stream (blues symbols in the
figure) represent the models studied in literature \citet[e.g][]{Sacco2008,
Sacco2010}. Both the observed and synthesized luminosities are
unabsorbed, that is we did not include any absorption due to the
interstellar medium. The X-ray emission
arising from impact regions is expected to undergo significant absorption
by the optically thick chromosphere in which the accretion column
is partially rooted (e.g. \citealt{Sacco2010, Curran2011A&A...526A.104C,
Reale2013, Bonito2014ApJ...795L..34B}). Our models do not take
into account this local absorption and, in fact, all of them (blue,
red, and green symbols in the figure) slightly overestimate the
luminosity of X-ray lines.

On the other hand, we note that models without radiative heating
(blue symbols) strongly underestimate (even by several orders of
magnitude) the luminosity of UV lines, thus failing to reproduce
the luminosity of UV emission lines. Models including the radiative
heating (green and red symbols in Fig.~\ref{fig:Lum.1}) predict
larger luminosity of UV lines that are closer to the observed ones.
In particular the observations are in between the two extreme
limiting cases considered, namely models either overestimating
(green symbols) or neglecting (red) the radiative cooling in the
unshocked stream. Analogous results were found for all the
models explored and characterized by different stream density and/or
metallicity. We want to stress the fact that the models 
RT and RT-PR predict a X-ray vs. UV lines luminosity ratio closer 
to the observed one respect to models without radiative heating,
independently from the normalization used. 
In the light of these results, we suggest that the
excess of UV emission observed in CTTSs and related to accretion
impacts originates from material of the unshocked stream heated
radiatively by the post-shock plasma at the base of the stream
itself.

\section{Summary and conclusions}
\label{sec.5}

We modelled the impact of an accretion stream onto the surface
of a CTTS and developed a method to investigate the effect
of radiative heating of the unshocked stream material by the
post-shock plasma at the base of the accretion stream. Our model
calculates self-consistently the heating feedback due to the 
absorption
by the accretion stream of the UV/soft-X-ray emission generated by
the shocked plasma of the impact region. Our main results are:

\begin{itemize}
\item The accretion stream is optically thick to the UV/soft X-ray
emission originating from the post-shock zone, and the absorption
heats up the downfalling plasma up to few $10^5\;$K before
the impact; we defined this pre-heated region precursor;

\item The precursor does not affect the characteristics of
the shock heated plasma; the X-ray spectra synthesized from models
either including or neglecting the radiative heating of the unshocked
stream material are almost the same;

\item We found that the precursor in the inflowing material
is an important UV emitter; models including the effect of radiative
heating predict luminosity of UV lines larger than that of models
without radiative heating; we explored two extreme limiting cases
in which the unshocked stream material is assumed to be either
optically thick (with no radiative cooling) or optically thin; we
found that the range of luminosities of UV lines derived from the
models cover the luminosities inferred from observations of three
selected CTTSs (namely MP~Mus, TW~Hya, and V4046~Sgr); we note that
models neglecting the radiative heating underestimate the UV emission
by several orders of magnitude.
\end{itemize}

Our study provides insight on the  effects 
of the absorption of X-ray radiation due to the downfalling
stream, on their consequences on the dynamics and energetics of 
the plasma at the base of the accretion column, and on the upper 
and lower limits of the range of such effects. In particular, our results 
underline the importance of
including the radiative transfer in models describing the accretion
phenomenon. This is a necessary step to determine accurately the
structure of the impact region and the radiation emerging from the
base of the accretion column and emitted in different bands. The
development of a radiation hydrodynamic model (thus including
self-consistently the radiative transfer in the calculation) would 
be the next natural step to address the issue of the optical thickness
of the pre-shock material. Some work has been done in this direction
(e.g. \citealt{Chieze2012, 2013A&A...549A.126I, 2014EPJWC..6404002D})
but a full detailed study of radiative transfer effects in accretion
impacts onto the surface of CTTSs is still missing. In the next
future we plan to address this issue by including self-consistently
the radiative transfer effects in our model.

\begin{acknowledgements}PLUTO is developed at the Turin
Astronomical Observatory in collaboration with the Department of
Physics of the Turin University. We acknowledge the HPC facility
(SCAN) of the INAF - Osservatorio Astronomico di Palermo, for having
provided high performance computing resources and support.
R. B. acknowledges financial support from INAF under PRIN2013 Programme
'Disks, jets and the dawn of planets'
\end{acknowledgements}

\bibliographystyle{aa}
\bibliography{biblio}

\begin{table*}
\caption{Observed luminosity of the selected lines in UV and X-ray bands of CTTSs.} 
\label{tab:Lum.rig.oss} 
\centering
\begin{tabular}{cccccc}
	\hline 
	\hline
   Line & $T_{max}$ & Ion & MP Mus & TW Hya & V4046 Sgr \\
$\lambda$ [\AA] & [MK] &   & $\rm [10^{40}\;ph\;s^{-1}]$ & $\rm [10^{40}\;ph\;s^{-1}]$ & $\rm [10^{40}\;ph\;s^{-1}]$ \\
	\hline
     12.13 & 6.3 & \ion{Ne}{X} & 0.0022 $\pm$ 0.0003 & 0.00301 $\pm$ 0.00007 & 0.0031 $\pm$ 0.0002\\
     13.45 & 4.0 & \ion{Ne}{IX} & 0.0022 $\pm$ 0.0004 & 0.00712 $\pm$ 0.00017 & 0.0035 $\pm$ 0.0002\\
     18.97 & 3.2 & \ion{O}{VIII} & 0.0080 $\pm$ 0.0006 & 0.0091 $\pm$ 0.0004 & 0.0055 $\pm$ 0.0003\\
     21.60 & 2.0 &\ion{O}{VII} & 0.0042 $\pm$ 0.0006 & 0.0053 $\pm$ 0.0005 & 0.0062 $\pm$ 0.0003\\
     24.78 & 2.0 & \ion{N}{VII} & 0.0014 $\pm$ 0.0004 & 0.0030 $\pm$ 0.0003 & 0.0071 $\pm$ 0.0004\\
     1238.80 & 0.20  & \ion{N}{V} & 0.616 $\pm$ 0.011 & 1.18 $\pm$ 0.02 & 1.724 $\pm$ 0.006\\
     1242.80 & 0.20 & \ion{N}{V} & 0.267 $\pm$ 0.011 & 0.527 $\pm$ 0.012 &  0.607 $\pm$ 0.006\\
     1393.80 & 0.08 & \ion{Si}{IV} & 0.90 $\pm$ 0.10 & 0.29 $\pm$ 0.10 & <0.06\\
     1402.80 & 0.08 & \ion{Si}{IV} & 0.51 $\pm$ 0.10 & 0.25 $\pm$ 0.10 & <0.06\\
     1548.20 & 0.13 & \ion{C}{IV} & 9.36 $\pm$ 0.10 & 16.05 $\pm$ 0.07 & 3.440 $\pm$ 0.006\\
	\hline 
	\end{tabular} 
\end{table*}

\begin{table*}
\caption{Luminosity of the selected lines in UV and X-ray bands derived from models with density $n_{acc} = 10^{11}$ cm$^{-3}$ (D1e11-A05 and D1e11-A08).} 
\label{tab:Lum.rig.mod} 
\centering
\begin{tabular}{ccccccccc}
	\hline 
	\hline

Line & $T_{max}$ & Ion & A05 & A05-RT & A05-RT-PR & A08 & A08-RT & A08-RT-PR\\
$\lambda$ [\AA] & [MK] &   & $\rm [10^{40}\;ph\;s^{-1}]$ & $\rm [10^{40}\;ph\;s^{-1}]$ & $\rm [10^{40}\;ph\;s^{-1}]$ & $\rm [10^{40}\;ph\;s^{-1}]$ & $\rm [10^{40}\;ph\;s^{-1}]$ & $\rm [10^{40}\;ph\;s^{-1}]$ \\
	\hline
12.1321 & 6.3 & \ion{Ne}{X} & 0.0163 & 0.0233 & 0.0164 & 0.0083 &    0.0150 & 0.0093\\
13.4471 & 4.0 & \ion{Ne}{IX}   &    0.0216 &    0.0280 &    0.0222 &   0.0125 &    0.0184 &   0.0132    \\
18.9671 & 3.2 & \ion{O}{VIII}  &    0.2092 &    0.2775 &    0.2172 &   0.1245 &    0.1812 &   0.1305    \\
21.6020 & 2.0 &\ion{O}{VII}    &    0.0973 &    0.1293 &    0.1030 &   0.0616 &    0.0874 &   0.0639    \\
24.7793 & 2.0 & \ion{N}{VII}   &    0.0223 &    0.0297 &    0.0234 &   0.0137 &    0.0198 &   0.0143    \\
1238.82 & 0.20  & \ion{N}{V}   &    0.2429 &   24.6802 &    0.2814 &   0.1223 &   19.7046 &   0.1680    \\
     1242.81 & 0.20 & \ion{N}{V}    &    0.1216 &   12.3678 &    0.1409 &   0.0613 &    9.8745 &   0.0841    \\
     1393.76 & 0.08 & \ion{Si}{IV}  &    0.0740 &   15.5503 &    0.5627 &   0.0342 &   14.1564 &   0.4197    \\
     1402.77 & 0.08 & \ion{Si}{IV}  &    0.0375 &    7.8720 &    0.2851 &   0.0173 &    7.1669 &   0.2129    \\
     1548.19 & 0.13 & \ion{C}{IV}   &    0.8114 &   212.931 &    3.0525 &   0.4349 &  182.858  &   0.9807    \\	
	\hline 
	\end{tabular} 
\end{table*}

\begin{table*}
\caption{Luminosity of the selected lines in UV and X-ray bands derived from models with density $n_{acc} = 5\times 10^{11}$ cm$^{-3}$ (D5e11-A05, D5e11-A08).} 
\label{tab:Lum.rig.mod2} 
\centering
\begin{tabular}{ccccccccc}
	\hline 
	\hline

Line & $T_{max}$ & Ion & A05 & A05-RT & A05-RT-PR & A08 & A08-RT & A08-RT-PR\\
$\lambda$ [\AA] & [MK] &   & $\rm [10^{40}\;ph\;s^{-1}]$ & $\rm [10^{40}\;ph\;s^{-1}]$ & $\rm [10^{40}\;ph\;s^{-1}]$ & $\rm [10^{40}\;ph\;s^{-1}]$ & $\rm [10^{40}\;ph\;s^{-1}]$ & $\rm [10^{40}\;ph\;s^{-1}]$ \\
	\hline
     12.1321 & 6.3  & \ion{Ne}{X}   &   0.0045 &   0.0093 &  0.0046 &   0.0052 &   0.0079 &  0.0052    \\
     13.4471 & 4.0  & \ion{Ne}{IX}  &   0.0081 &   0.0139 &  0.0084 &   0.0100 &   0.0107 &  0.0081    \\
     18.9671 & 3.2  & \ion{O}{VIII} &   0.0825 &   0.1354 &  0.0854 &   0.1030 &   0.1034 &  0.0810    \\
     21.6020 & 2.0  &\ion{O}{VII}   &   0.0430 &   0.0632 &  0.0445 &   0.0536 &   0.0480 &  0.0410    \\
     24.7793 & 2.0  & \ion{N}{VII}  &   0.0092 &   0.0144 &  0.0096 &   0.0115 &   0.0110 &  0.0090    \\
     1238.82 & 0.20 & \ion{N}{V}    &   0.0496 &  17.6187 &  0.1006 &   0.0858 &  27.2675 &  0.1926    \\
     1242.81 & 0.20 & \ion{N}{V}    &   0.0248 &   8.8297 &  0.0504 &   0.0430 &  13.6649 &  0.0965    \\
     1393.76 & 0.08 & \ion{Si}{IV}  &   0.0503 &  15.9665 &  0.1064 &   0.0590 &  22.9215 &  0.2190    \\
     1402.77 & 0.08 & \ion{Si}{IV}  &   0.0255 &   8.0848 &  0.0540 &   0.0299 &  11.6061 &  0.1110    \\
     1548.19 & 0.13 & \ion{C}{IV}   &   0.2119 & 187.321  &  0.3115 &   0.3154 & 275.393  &  1.6222    \\
	\hline 
	\end{tabular} 
\end{table*}

\end{document}